\documentclass{article} 
\usepackage{iclr2026_conference,times}

\usepackage{amsmath,amsfonts,bm}

\def\eqref#1{equation~\ref{#1}}

\def\1{\bm{1}}

\DeclareMathAlphabet{\mathsfit}{\encodingdefault}{\sfdefault}{m}{sl}
\SetMathAlphabet{\mathsfit}{bold}{\encodingdefault}{\sfdefault}{bx}{n}

\usepackage{hyperref}
\usepackage{url}

\usepackage{amsmath,amssymb,mathtools}

\usepackage{algorithm}
\usepackage{algpseudocode}
\usepackage{xcolor}
\usepackage{graphicx}
\usepackage[most]{tcolorbox}
\tcbuselibrary{breakable}
\usepackage{enumitem}
\usepackage{tabularx}
\usepackage{wrapfig}
\usepackage{booktabs}

\AtBeginDocument{\let\cite\citep}

\title{Beyond Better Ideas: How Human–AI Collaboration Shapes Scientific Ideas and Human Perceptions}

\author{Qian Chen$^{1}$\thanks{Qian Chen and Saatvik Pradhan contributed equally to this work and share first authorship.}
\quad Saatvik Pradhan$^{1}$\footnotemark[1]
\quad Janice Ahn$^{1}$
\quad Berk Atil$^{1}$
\quad Angkai Li$^{2}$ \\
\textbf{Shijie Lu$^{3}$
\quad Nobin Sarwar$^{4}$
\quad Michael Ulis$^{1}$
\quad Ruyuan Wan$^{1}$
\quad Wenpeng Yin$^{1}$} \\
$^{1}$The Pennsylvania State University
\quad $^{2}$Harvard University
\quad $^{3}$Arizona State University \\
$^{4}$University of Maryland, Baltimore County \\
\texttt{\{quc20,szp6118,wenpeng\}@psu.edu}
}

\iclrfinalcopy 
\begin{document}
\raggedbottom

\maketitle
\lhead{}
\renewcommand{\headrulewidth}{0pt}

\begin{abstract}
As AI increasingly becomes a collaborator in scientific research, its impact may extend beyond improving ideas to reshaping how they are evaluated, attributed, and perceived by humans. Yet existing research has largely examined these dimensions in isolation. We present a \textit{unified empirical study of Human–AI scientific ideation}, investigating two complementary perspectives: i) \textit{how AI collaboration shapes scientific ideas and their evaluation}, and ii) \textit{how firsthand collaboration reshapes human perceptions of AI}. Through a shared experimental setting involving Human–AI and Human–Human research teams, human and LLM evaluators, collaboration-provenance analysis, and participant surveys, we uncover several counterintuitive findings. First, AI collaboration shows no significant advantage in human-evaluated idea quality, while iterative revision consistently improves quality. Second, AI judges can approximate human scores yet disagree on idea rankings; notably, an AI judge favors AI-assisted proposals without a corresponding advantage in human evaluations. Third, collaboration provenance remains only partially detectable, revealing substantial overlap between Human–AI and Human–Human scientific outputs. Finally, firsthand AI collaboration is associated with greater perceived usefulness but smaller efficiency gains than expected, suggesting a nuanced recalibration of human expectations. Overall, our findings challenge the notion that AI's value in scientific ideation lies primarily in producing better ideas, revealing instead its broader influence on \textit{scientific creation, evaluation, attribution, and human perceptions}.
\end{abstract}

\section{Introduction}

Large language models (LLMs) are rapidly moving from tools that merely retrieve or summarize scientific knowledge to active participants in the production of new knowledge. Researchers increasingly use generative AI to brainstorm hypotheses, formulate research questions, design experiments, and refine scientific arguments, while LLMs are simultaneously being explored as evaluators of research ideas, proposals, and scientific outputs. This convergence creates a consequential new regime for science: \emph{AI may increasingly participate on both sides of the scientific process---helping to generate ideas and helping to decide which ideas are valuable}. Understanding this regime is urgent. If AI changes what researchers propose, how those proposals are evaluated, or even researchers' own beliefs about what constitutes good scientific reasoning, its influence can propagate far beyond individual interactions to shape the allocation of attention, resources, and ultimately the direction of scientific discovery. The central question is therefore no longer simply whether AI can generate plausible research ideas, but \emph{how scientific ideation changes when AI becomes simultaneously a collaborator, an evaluator, and an increasingly difficult-to-observe contributor}.

Despite growing interest in AI-assisted science, these roles have largely been studied in isolation. Prior work has examined whether LLMs can generate novel hypotheses or research ideas~\cite{DBLPSiYH25,zhou-etal-2024-hypothesis}, whether Human--AI collaboration improves creative or knowledge-intensive work~\cite{noy2023productivity}, and whether LLMs can approximate human judgments as automated evaluators or ``AI judges''~\cite{DBLPhengC00WZL0LXZ23,DBLPWangLCCZLCKLLS24}. Yet these research threads leave several fundamental questions unanswered. \emph{Does collaborating with AI actually produce research ideas that human experts consider better?} If AI itself evaluates those ideas, \emph{does it preserve human preferences, or does it systematically value AI-assisted work differently?} As AI-generated and human-generated reasoning become increasingly intertwined, \emph{does the resulting methodology retain detectable traces of human intellectual participation?} And beyond the artifacts themselves, \emph{does firsthand collaboration recalibrate researchers' beliefs about AI's scientific capabilities?} These questions cannot be answered by studying generation, collaboration, or evaluation independently. They require observing the entire pipeline---from how research ideas are produced, to how humans and AI evaluate them, to what their provenance reveals, and finally to how collaboration changes the humans involved.

To enable such an investigation, we construct a controlled empirical testbed centered on scientific research proposals. Our real-world corpus contains 70 one-page NLP/ML research proposals produced by 35 student teams under Human+AI and Human+Human collaboration conditions, with both initial and revised submissions spanning 20 research questions. Each proposal is evaluated by human reviewers and multiple state-of-the-art LLM judges using the same rubric, covering novelty, feasibility, detail and clarity, broad applicability, correctness and soundness, and overall quality, together with within-question rankings. To study whether collaboration provenance can be inferred beyond surface stylistic cues, we additionally construct a synthetic training corpus derived from 1,000 human-authored research papers, creating paired methodologies that differ in the information available during their construction, and evaluate transfer exclusively on the real student proposals. Finally, pre- and post-study questionnaires allow us to examine how direct collaboration changes participants' perceptions of GenAI. This design connects \emph{creation, evaluation, provenance, and human perception} within a single empirical framework rather than treating them as separate phenomena.

Our results reveal a considerably more nuanced picture than either ``AI improves science'' or ``AI fails to match humans.'' First, Human+AI and Human+Human teams achieve nearly indistinguishable overall quality under human evaluation; both improve substantially through revision, suggesting that the dominant benefit in our setting comes from iterative refinement rather than AI participation itself. Second, AI judges can appear surprisingly human-like under absolute scoring while still disagreeing with humans about \emph{which proposal is better}: GPT-4.1 achieves the strongest score calibration among the tested models, yet a substantial gap remains in preference ranking, revealing that \emph{score calibration and preference alignment are distinct properties of an evaluator}. Third, and more strikingly, evaluation is not neutral to collaboration provenance. GPT-4.1 assigns systematically higher scores to Human+AI proposals even when the human panel shows essentially no corresponding advantage, with the largest divergence appearing in ``broad applicability''. Thus, introducing AI on both sides of the pipeline can create an \emph{evaluator-side amplification effect}: AI may make AI-assisted work appear better without an equivalent shift in human judgment. Fourth, collaboration leaves a detectable but weak provenance signal: synthetic supervision transfers to real proposals and enables above-chance identification of collaboration condition, but accuracy remains only about 66--67\%, indicating substantial overlap between methodologies produced with and without the corresponding human contribution. Finally, firsthand Human--AI collaboration does not simply make participants more optimistic about AI. Instead, it \emph{recalibrates} their mental model: participants perceive greater practical helpfulness and capability while simultaneously reporting less speed-up and slightly lower confidence in integrating GenAI than they initially expected. Together, these findings suggest that AI's impact on scientific ideation is not a single capability gain, but a coupled transformation of \emph{how ideas are produced, judged, attributed, and understood by humans}.

Our work makes the following contributions:
\begin{itemize}[leftmargin=*]
    \item \textbf{An end-to-end empirical framework for AI-mediated scientific ideation.}
    We jointly study AI as a collaborator, AI as an evaluator, collaboration provenance, and researchers' perception shifts, connecting stages of the scientific ideation pipeline that prior work has largely examined independently.

    \item \textbf{A controlled proposal corpus linking collaboration to human and AI evaluation.}
    We construct a dataset of 70 real research proposals from Human+AI and Human+Human teams, including revision trajectories, multi-dimensional human judgments, within-question rankings, and evaluations from multiple frontier LLMs, complemented by a synthetic corpus for studying provenance transfer.

\item \textbf{A broader account of AI evaluation beyond human-score approximation.}
We show that reliable AI evaluation cannot be reduced to reproducing human scores: an AI judge may be well calibrated in absolute scores while diverging from human preference orderings and, more consequentially, may systematically favor AI-assisted work even when human reviewers do not. These findings reveal AI evaluation as an active component of the Human--AI scientific ecosystem, where \emph{how knowledge is produced} and \emph{who evaluates it} can interact to shape conclusions about scientific quality.

    \item \textbf{A broader account of what Human--AI collaboration changes.}
    Beyond proposal quality, we show that collaboration provenance remains partially detectable and that direct experience with GenAI recalibrates researchers' expectations in multiple directions, providing evidence that AI changes not only scientific artifacts but also the human judgments and beliefs surrounding them.
\end{itemize}

\section{Related Work}
\label{sec:related}

Our work connects several lines of research that have largely developed
separately: Human--AI collaboration, AI-assisted scientific ideation,
LLM-based evaluation, collaboration provenance, and human perceptions
of AI.

\paragraph{\textbf{Human--AI Collaboration and Scientific Ideation.}}
Prior work has established LLMs as active collaborators in creative and
knowledge-intensive tasks. CoAuthor~\cite{lee2022coauthor} studied
human--LM interaction during writing, while controlled experiments show
that generative AI can improve productivity and output
quality~\cite{noy2023productivity}, albeit with important tradeoffs.
AI assistance can increase individual creativity while reducing
collective diversity~\cite{doshi2024creativity}, and its benefits can
reverse when tasks fall outside the model's capability
frontier~\cite{dellacqua2026jagged}. Together, these results suggest
that Human--AI collaboration should be evaluated across multiple
dimensions rather than through a single measure of quality.

This question is particularly consequential in scientific ideation. Chen et al.~\cite{chen2026philosophersstone} find that AI use in scientific
publications is associated with higher novelty and disruption on average, with
differences across levels of creativity, fields, and types of AI tools. Systems such as SciMON~\cite{wang2024scimon} explicitly optimize
scientific ideas for novelty, while increasingly autonomous systems
such as the AI Scientist~\cite{lu2024aiscientist} automate substantial
parts of the research workflow. Most closely related, Si et
al.~\cite{DBLPSiYH25} find that LLM-generated research ideas can
be judged more novel but less feasible than human-generated ones.
Whereas this literature primarily contrasts \emph{human-generated}
and \emph{AI-generated} research, we study the intermediate and
increasingly common regime of \emph{Human--AI collaboration}: humans
remain responsible for research design while AI participates in
developing and revising the ideas.

\paragraph{\textbf{LLMs as Evaluators and Scientific Judges.}}
A parallel literature asks whether LLMs can approximate human
judgments as scalable evaluators. Studies such as
G-Eval~\cite{liu2023geval} and MT-Bench/Chatbot
Arena~\cite{DBLPhengC00WZL0LXZ23} demonstrate substantial alignment
between LLM and human evaluations, motivating broader use of
LLM-as-a-Judge methods~\cite{chiang2023alternative,
bavaresco2025judgebench}. Yet agreement varies substantially across
tasks, criteria, and evaluation protocols~\cite{thakur2024judgingjudges},
and LLM judges exhibit systematic biases even when aggregate
human--AI agreement appears strong~\cite{gu2024judgesurvey}.

Scientific evaluation makes this problem especially challenging:
novelty, feasibility, and soundness are inherently multidimensional,
and even expert reviewers exhibit substantial
disagreement~\cite{cortes2021inconsistency,beygelzimer2023arbitrary}.
Indeed, GPT-generated scientific feedback can overlap with human
feedback at levels approaching human--human
agreement~\cite{liang2023feedback}. We therefore use agreement among human reviewers as a benchmark for evaluating AI judges and, importantly,
distinguish \emph{score calibration} from \emph{preference alignment}:
assigning human-like scores does not necessarily imply ranking
scientific ideas as humans do.

\paragraph{\textbf{Creator--Evaluator Interaction.}}

Existing research has identified position, verbosity, and other biases
in LLM judges~\cite{DBLPhengC00WZL0LXZ23}, including evidence that models
can recognize and preferentially evaluate their own
generations~\cite{panickssery2024selfpreference}. We extend this
question from properties of the generated text to the
\emph{process that produced it}. Specifically, we ask whether an AI
judge evaluates work created through Human--AI collaboration
differently from comparable Human+Human work, even when human
evaluators show no such preference.

This motivates a broader \emph{creator--evaluator interaction}:
measured gains from AI-assisted science may depend not only on the
quality of the resulting work, but also on who evaluates it. This
interaction becomes increasingly important as AI participates on both
the creation and evaluation sides of the scientific pipeline.

\paragraph{\textbf{Detecting Collaboration Provenance.}}

AI-generated text detection has proven fragile: paraphrasing can
substantially degrade detectors~\cite{sadasivan2024detection,
krishna2023paraphrasing}, while detectors may capture unintended
linguistic characteristics such as non-native writing
style~\cite{liang2023detectorbias}. Related work has therefore also
considered corpus-level estimation of AI-modified
content~\cite{liang2024monitoring}.

Our provenance problem is fundamentally different from conventional
human-vs.-AI detection. Both of our conditions involve human authors;
we instead ask whether the \emph{collaboration process}---Human+AI
versus Human+Human---leaves detectable signatures in the resulting
scientific methodology. This shifts provenance analysis from
identifying who wrote a text to inferring how an intellectual artifact
was produced.

\paragraph{\textbf{Human Perceptions of AI Collaboration.}}

Human--AI teaming also depends on how people perceive their AI
partners. Prior work shows that expectations and pre-existing
attitudes shape willingness to collaborate with
AI~\cite{zhang2021idealhuman}, while educational studies highlight the
role of norms and disclosure expectations in students' use of
generative AI~\cite{adnin2025undisclosed}. We complement these studies
by examining perceptions around a concrete scientific collaboration
experience, asking whether hands-on Human--AI ideation recalibrates
participants' views of AI's usefulness, trustworthiness, efficiency,
and role in research.
\section{Data Collection}
\label{sec:data-collection}

Our study integrates two complementary types of data to capture both
the \emph{outputs} and the \emph{human experience} of Human--AI
scientific collaboration. First, our \textbf{scientific ideation data}
combine a large-scale synthetic corpus for model training with a
controlled real-world corpus of research proposals for evaluation.
The real-world proposals are further annotated by both human reviewers
and AI judges using the same multidimensional scientific-quality
rubric. Second, our \textbf{participant questionnaire data} capture
participants' expectations and perceptions of generative AI before
and after the scientific-ideation study. Jointly, these data allow us
to study Human--AI scientific collaboration from three complementary
perspectives: the artifacts it produces, how those artifacts are
evaluated, and how the experience changes the humans involved.

\subsection{Scientific Ideation Data}
\label{sec:ideation-data}

\paragraph{\textbf{Synthetic Training Corpus.}}
To obtain scalable supervision without using any real participant data
for training, we construct a synthetic corpus from 1,000
human-authored research papers. Each paper provides a research
question together with its original methodology, allowing us to
construct two controlled methodology-generation conditions.

For each paper, GPT-4.1 generates two methodology descriptions. In
the \emph{human-grounded} condition, the model receives the research
question (\emph{RQ}) together with information from the source paper, grounding
its methodology in an existing human-designed solution. In the
\emph{RQ-only} condition, GPT-4.1 receives only the research question
and independently generates a methodology. We use these conditions as
synthetic proxies for Human+Human and Human+AI methodological
development, respectively. Importantly, GPT-4.1 verbalizes both
conditions; the controlled distinction is therefore the
\emph{source of methodological reasoning}, rather than surface
authorship. This procedure yields 2,000 methodology instances---1,000 per
condition.

\paragraph{\textbf{Real-World Proposal Corpus as Testing Data.}}

The real-world corpus comes from a controlled Human--AI scientific
ideation study in which participants worked in teams to develop
one-page research proposals addressing predefined NLP/ML research
questions. Teams operated under either a \emph{Human+Human} or
\emph{Human+AI} collaboration condition.

The corpus contains 70 proposal identities---36 Human+AI and 34
Human+Human---covering 20 research questions, listed in
Appendix Table~\ref{tab:proposal-prompts}.   The study
spans both undergraduate and graduate cohorts. Each team submitted an
initial proposal (v1) and a revised proposal (v2) one week later,
providing a longitudinal view of how scientific ideas evolve through
revision.

Then, those real-world proposals are evaluated independently by human
reviewers and AI judges using an identical scientific-quality rubric.
Evaluators assign integer scores from 1 (\emph{very poor}) to 5
(\emph{outstanding}) for five dimensions:
\emph{novelty}, \emph{feasibility}, \emph{detail and clarity},
\emph{broad applicability}, and \emph{correctness and soundness},
together with an \emph{overall quality} score. Each assessment is
accompanied by a short written justification. Evaluators additionally rank proposals addressing the same research
question from strongest to weakest. The annotations therefore provide
two complementary views of scientific judgment: \emph{absolute
quality}, captured by multidimensional scores, and \emph{relative
preference}, captured by within-RQ rankings. This common protocol
enables direct comparison of human and AI evaluation.

No real-world proposal is used to train or fine-tune the provenance
classifier. The real corpus therefore provides a strict
synthetic-to-real transfer evaluation.

\subsection{Participant Questionnaire Data}
\label{sec:questionnaire-data}

\begin{wraptable}[11]{r}{0.45\textwidth}
\vspace{-22pt}
  \centering
  \caption{Questionnaire coverage and collaboration-condition mapping.}
  \label{tab:rq5-coverage}
  \scriptsize
  \renewcommand{\arraystretch}{1.08}

  \begin{tabularx}{\linewidth}{@{}Xr@{}}
    \toprule
    Result & Count \\
    \midrule
    Unique pre-survey participants & 47 \\
    Unique post-survey participants & 45 \\
    Present in both surveys & 37 \\
    Pre-survey only & 10 \\
    Post-survey only & 8 \\
    Unique participants overall & 55 \\
    Assigned Human+AI & 19 \\
    Assigned Human+Human & 31 \\
    Still unmatched & 5 \\
    \bottomrule
  \end{tabularx}
\end{wraptable}
In addition to the scientific artifacts and their evaluations, we
collect questionnaire data to characterize how direct collaboration
with generative AI changes participants' perceptions of its role in
scientific research. Participants completed questionnaires before
the study and after completing the research-design task. The
pre-study questionnaire captures expectations about GenAI, whereas
the post-study questionnaire captures perceptions after participants
had experienced the collaboration process.

We identify eight constructs that appear in comparable form in both
questionnaires:
(i) confidence in integrating GenAI into research,
(ii) overall attitude toward GenAI,
(iii) perceived helpfulness across research-design tasks,
(iv) perceived severity of challenges associated with GenAI,
(v) expected or experienced speed-up of research ideation,
(vi) human effort required to refine GenAI outputs,
(vii) perceived enhancement of research-design capabilities, and
(viii) trust in GenAI recommendations.
Where the pre- and post-study questionnaires used different response
encodings, we harmonize semantically equivalent responses onto a
common 1--5 ordinal scale, preserving the direction of each construct.

The questionnaire data contain pre-study responses from 47
unique participants and post-study responses from 45 unique
participants. Thirty-seven participants appear in both surveys.
Across the study, 55 unique participants can be identified, including
19 assigned to Human+AI, 31 to Human+Human, and five whose
collaboration condition could not be reliably matched. Because our
perception analysis concerns expectations about versus experiences
with GenAI collaboration, the corresponding pre/post analysis focuses
on participants assigned to the Human+AI condition.

\section{Experiments}
\label{sec:research-questions}

We investigate five interconnected research questions to understand how scientific ideation changes when AI serves not only as a collaborator in developing research ideas but also as an evaluator of their quality. Rather than examining AI generation and evaluation in isolation, we study their interactions across the scientific ideation pipeline. We first ask whether Human+AI collaboration produces research proposals that human experts consider better, and how other factors shape proposal quality (\textbf{RQ1}). We then examine whether AI judges can faithfully reproduce human assessments, distinguishing absolute score agreement from relative preference alignment (\textbf{RQ2}). Crucially, we investigate whether AI judges systematically favor AI-assisted proposals even when human reviewers do not, revealing potential interactions between how ideas are produced and how they are evaluated (\textbf{RQ3}). Beyond evaluation, we ask whether collaboration leaves detectable traces in the resulting proposals (\textbf{RQ4}), and whether firsthand Human+AI collaboration reshapes researchers' perceptions of AI's scientific capabilities (\textbf{RQ5}). Jointly, these questions move beyond whether AI improves individual research tasks to examine how its simultaneous roles as collaborator, evaluator, and contributor may reshape the production, assessment, attribution, and human perception of GenAI.

\subsection{RQ1: How do different factors (e.g., collaboration style, proposal revision, researcher experience, and task difficulty) affect the quality of scientific research proposals?}
\label{sec:rq1}

\textbf{Motivation.}
As AI becomes increasingly involved in scientific ideation, a fundamental question is whether AI collaboration actually improves the quality of research ideas, and how its contribution compares with other factors shaping scientific creativity. Beyond collaboration style, proposal quality may depend on iterative refinement, researchers' experience, and the difficulty of the research problem. Understanding their relative roles is essential for distinguishing the benefits of AI assistance from those of established research practices.

\textbf{Method.}
We first compare Human+AI and Human+Human proposals across both submission
rounds. This comparison tests whether reviewers assigned different overall
scores or quality profiles to the two collaboration conditions. Human+AI and
Human+Human proposals addressed the same research question in only one of the
20 cases. We therefore interpret the condition estimates as associations in
this sample rather than effects caused solely by AI collaboration.

We next study how each team's proposal changed from v1 to v2. Each team
continued working on the same research problem, so the within-team comparison
directly captures changes made during the one-week revision period. Finally,
we compare the gains of the two conditions using the condition-by-round
interaction:
\[
  \Delta\Delta =
  (\mathrm{v2}-\mathrm{v1})_{\mathrm{Human+AI}}
  -
  (\mathrm{v2}-\mathrm{v1})_{\mathrm{Human+Human}}.
\]
A positive $\Delta\Delta$ indicates a larger revision gain for Human+AI; a
negative value indicates a larger gain for Human+Human.

Each observation is one reviewer's score for one proposal. Each team
contributed both v1 and v2. For each research question, reviewers scored and
ranked all proposals submitted by the different teams. We use a linear
mixed model so that these repeated observations do not count as independent
evidence. The model accounts for consistent differences among reviewers and
teams when estimating the collaboration-condition and revision effects.

We use overall quality and normalized rank as the primary outcomes and analyze
the five rubric dimensions as exploratory outcomes. For each comparison, this
produces five dimension-level tests. Testing several dimensions increases the
chance that one will appear significant by chance. We therefore adjust each
set of five $p$-values using the Benjamini--Hochberg false-discovery-rate
procedure; $q$ denotes the adjusted $p$-value. For v1--v2 changes, we report
$d_z$, which expresses the mean within-team change in standard-deviation
units. For collaboration-condition differences, we report Hedges' $g$, which
expresses the between-condition difference in standard-deviation units and
includes a small-sample correction.

\textbf{Results.} We examine how four factors---collaboration style, proposal revision, researcher experience, and task difficulty---relate to human-evaluated proposal quality.

\textbullet\enspace\textbf{Collaboration style: AI assistance does not significantly improve human-evaluated proposal quality.}
We compare the two collaboration conditions using each proposal's average
overall quality score across human reviewers. The results show little difference.
Averaged across v1 and v2, Human+AI proposals scored 0.04 points higher than
Human+Human proposals (95\% CI $[-0.31,0.40]$, Hedges' $g=0.08$). The small
effect size and the confidence interval spanning zero indicate similar overall
ratings in the two conditions.

Although the overall ratings were similar, the dimension-level mean
differences pointed in different directions, as shown in Table~\ref{tab:rq1-profile}. Human+AI proposals scored higher on correctness and soundness ($+0.25$) and novelty ($+0.19$). Human+Human proposals scored higher on feasibility ($+0.22$), broad applicability ($+0.16$), and detail and clarity ($+0.11$). All five confidence intervals include zero. We therefore treat this pattern as a direction for future study rather than evidence of dimension-level condition effects.

\begin{table*}[htbp]
  \centering
  \small
  \caption{Mean contrasts for collaboration condition and revision. Values are
  rubric-point differences with 95\% confidence intervals. Positive condition
  values favor Human+AI; positive revision values favor v2. Lower normalized
  rank indicates a stronger proposal. The condition contrast for normalized
  rank is not applicable because the two conditions were generally ranked in
  different research-question groups.}
  \label{tab:rq1-profile}
  \begin{tabular}{lcc}
    \toprule
    Outcome & Human+AI $-$ Human+Human & v2 $-$ v1 \\
    \midrule
    Novelty & $+0.19$ $[-0.19,+0.58]$ & $+0.21$ $[+0.05,+0.37]$ \\
    Feasibility & $-0.22$ $[-0.57,+0.14]$ & $+0.03$ $[-0.13,+0.19]$ \\
    Detail \& clarity & $-0.11$ $[-0.43,+0.21]$ & $+0.10$ $[-0.06,+0.27]$ \\
    Broad applicability & $-0.16$ $[-0.48,+0.16]$ & $+0.04$ $[-0.03,+0.10]$ \\
    Correctness / soundness & $+0.25$ $[-0.06,+0.57]$ & $+0.07$ $[-0.00,+0.13]$ \\
    Overall quality & $+0.04$ $[-0.31,+0.40]$ & $+0.18$ $[+0.07,+0.30]$ \\
    Normalized rank & N/A & $-0.13$ $[-0.22,-0.05]$ \\
    \bottomrule
  \end{tabular}
\end{table*}

\textbullet\enspace\textbf{Proposal revision: Iterative refinement improves quality regardless of collaboration style.}
 Revision improved proposal quality overall, but the amount of improvement did
not differ significantly between Human+AI and Human+Human teams. Across all
teams, overall quality increased by 0.18 points (95\% CI $[0.07,0.30]$,
$d_z=0.54$,
$p=.003$). Normalized rank also improved from 0.56 to 0.43
($\Delta=-0.13$, 95\% CI $[-0.22,-0.05]$, $d_z=-0.53$, $p=.004$). Because a
lower rank indicates a stronger proposal, both outcomes show that reviewers
evaluated v2 more favorably than v1.

Table~\ref{tab:rq1-overall-trajectory} breaks this improvement down by
collaboration condition. Human+AI proposals increased from 3.39 at v1 to 3.52
at v2, a gain of 0.13 points. Human+Human proposals increased from 3.29 to
3.53, a gain of 0.24 points. Thus, the 0.10-point Human+AI advantage at v1
narrowed to a 0.01-point Human+Human advantage at v2.

\begin{table}[htbp]
  \centering
  \small
  \caption{Mean overall human rating by collaboration condition and submission
  round. Change equals v2 minus v1. The final row reports the between-condition
  gap at each round and the change in that gap.}
  \label{tab:rq1-overall-trajectory}
  \begin{tabular}{lrrr}
    \toprule
    Collaboration condition & v1 mean & v2 mean & Change \\
    \midrule
    Human+AI & 3.39 & 3.52 & $+0.13$ \\
    Human+Human & 3.29 & 3.53 & $+0.24$ \\
    \midrule
    Human+AI $-$ Human+Human & $+0.10$ & $-0.01$ & $-0.11$ \\
    \bottomrule
  \end{tabular}
\end{table}

Human+Human teams showed a 0.11-point larger numerical gain, but the
condition-by-round interaction was not statistically significant
($\Delta\Delta=-0.11$, $p=.58$). The data therefore support a general revision
benefit shared by both conditions, not a condition-specific difference in how
much proposals improved.

To identify which aspects of proposal quality changed during revision, we next
examine the five rubric dimensions. Revision increased novelty by 0.21 points
(95\% CI $[0.05,0.37]$, $d_z=0.44$, $p=.013$). The FDR-adjusted result was $q=.064$. Changes in the
other dimensions were smaller: detail and clarity increased by 0.10 points,
correctness and soundness by 0.07, broad applicability by 0.04, and feasibility
by 0.03. Their confidence intervals included zero. Revision therefore changed
perceived novelty and overall quality more than the dimensions tied to
execution.

\textbullet\enspace\textbf{Researcher experience: Higher academic level does not necessarily translate into higher proposal quality.}
Undergraduate proposals receive higher broad-applicability scores ($+0.40$, $q=.025$) and better normalized rankings ($-0.27$, $p=.024$) than graduate proposals addressing the same questions. This counterintuitive pattern suggests that academic level alone is not a reliable indicator of proposal quality. However, differences in cohort composition, research experience, and language background may contribute, preventing a causal interpretation.

\textbullet\enspace\textbf{Task difficulty: No detectable relationship with proposal quality.}
We also examined whether proposal ratings and within-question ranks differed
by course level or by the preassigned difficulty of the research question.
Undergraduate proposals received higher broad-applicability scores
($+0.40$, $q=.025$) and stronger normalized ranks ($-0.27$, $p=.024$) than
graduate proposals addressing the same questions. Differences in cohort
composition, prior research experience, or language background may contribute
to this pattern, but the present data do not identify its cause. Task difficulty
showed no detectable relationship with the proposal ratings. Because the study
included only 20 research questions across three difficulty levels, we
interpret this result as an absence of a large difficulty difference rather
than evidence that difficulty never matters.

\textbf{Conclusions.}
Our findings reveal that \textbf{proposal revision is more consistently associated with quality improvement than collaboration style}. Human+AI collaboration does not significantly outperform Human+Human collaboration under human evaluation, although the two settings exhibit suggestive differences across quality dimensions. Undergraduate proposals outperform graduate proposals on selected measures, while task difficulty shows no detectable association. Together, these results challenge the assumption that AI assistance or greater academic experience necessarily produces better scientific ideas, and highlight the importance of iterative refinement and multidimensional human evaluation.
\subsection{RQ2: How Well Does AI-as-Judge Align with Human Judgment?}
\label{sec:rq2}

\paragraph{\textbf{Motivation.}}
A key question for AI-assisted scientific evaluation is whether LLM judges can reliably evaluate research proposals at a level comparable to human reviewers, enabling them to assist or partially replace human evaluation.
To answer this question, we compare AI judgments against the aggregated human panel and further benchmark them against the agreement between individual human reviewers and the panel.
This reviewer--panel agreement serves as a reference for assessing whether AI judgments approach human-level consistency.

\paragraph{\textbf{Method \& Evaluation Metrics.}} We evaluate four representative LLM judges on the 70 proposals:
GPT-4.1~\cite{openai2025gpt41},
Gemini~2.5 Flash~\cite{google2025gemini25},
Claude Sonnet~4.6~\cite{anthropic2026sonnet46},
and Qwen~2.5 72B Instruct~\cite{qwen2024qwen25}. Each model follows the same two-stage evaluation pipeline.
First, every proposal is independently scored on the six evaluation dimensions using a 1--5 scale.
Second, proposals belonging to the same research question are ranked according to their predicted quality without re-scoring. 
The following templates summarize the prompts used for proposal
scoring and within-question ranking. The same templates were used
for GPT-4.1, Gemini~2.5 Flash, Claude Sonnet~4.6, and
Qwen~2.5 72B Instruct.

\begin{tcolorbox}[
  colback=gray!3,
  colframe=black!40,
  title={Prompt 1: Proposal Scoring},
  fonttitle=\bfseries,
  breakable,
  boxsep=2pt,
  left=4pt,
  right=4pt,
  top=4pt,
  bottom=4pt
]
\footnotesize
You are reviewing a one-page research proposal.

This rubric is designed to ensure consistent, fair, and
constructive evaluation of one-page research proposals.

Each proposal should be assessed on five key dimensions:
\textbf{Novelty}, \textbf{Feasibility},
\textbf{Detail \& Clarity}, \textbf{Broad Applicability Scope},
and \textbf{Correctness/Soundness}.
Reviewers assign a score from 1 (very poor) to 5 (outstanding)
for each dimension, accompanied by 2--3 sentences of comments.
An overall score should then be provided.

\textbf{Evaluation dimensions.}
\begin{itemize}[leftmargin=1.2em,itemsep=1pt,topsep=2pt]
  \item \textbf{Novelty:}
  1 = no novelty;
  5 = genuinely original or groundbreaking idea.

  \item \textbf{Feasibility:}
  1 = unfeasible under realistic constraints;
  5 = highly feasible with current resources, data, and technology.

  \item \textbf{Detail \& Clarity:}
  1 = incoherent or unclear;
  5 = excellent clarity and sufficient detail to allow reproduction.

  \item \textbf{Broad Applicability Scope:}
  1 = extremely narrow;
  5 = strong potential to generalize across domains and problems.

  \item \textbf{Correctness/Soundness:}
  1 = fundamentally flawed;
  5 = strong logical/mathematical foundation with no major flaws.
\end{itemize}

Evaluate the proposal naturally and thoughtfully as a human
reviewer would.
Be fair but critical.
Do not overinflate scores.
Base your evaluation only on the proposal content.

Return only JSON with:
\begin{itemize}[leftmargin=1.2em,itemsep=1pt,topsep=2pt]
  \item proposal ID;
  \item score and comments for Novelty, Feasibility,
  Detail\_Clarity, Broad\_Applicability\_Scope,
  and Correctness\_Soundness;
  \item Overall\_Rating score and comments.
\end{itemize}

\textbf{Input:} Proposal text.
\end{tcolorbox}

\begin{tcolorbox}[
  colback=gray!3,
  colframe=black!40,
  title={Prompt 2: Within-RQ Ranking},
  fonttitle=\bfseries,
  breakable,
  boxsep=2pt,
  left=4pt,
  right=4pt,
  top=4pt,
  bottom=4pt
]
\footnotesize
You are given research proposals written for the same research
question, along with their rubric-based evaluations.

Rank the proposals from strongest to weakest.

Rank~1 = strongest proposal.

Base the ranking on overall proposal quality, considering:
\begin{itemize}[leftmargin=1.2em,itemsep=1pt,topsep=2pt]
  \item Novelty
  \item Feasibility
  \item Detail \& Clarity
  \item Broad Applicability Scope
  \item Correctness/Soundness
\end{itemize}

Rank proposals naturally and thoughtfully as a human reviewer would.
Be fair but critical.
Do not overinflate rankings.
Base the ranking only on the proposal content and evaluation
results provided.
Do not re-score the proposals.

Return only JSON with a list of proposal rankings:
\begin{itemize}[leftmargin=1.2em,itemsep=1pt,topsep=2pt]
  \item rank;
  \item proposal\_id;
  \item short reason.
\end{itemize}

\textbf{Input:} Proposals and Step~1 evaluations.
\end{tcolorbox}

\paragraph{Evaluation Metrics.} For each proposal, we average human reviewers' scores separately for each
criterion and their ranks within the corresponding research question
(Section~\ref{sec:rq1}). We compare AI scores and rankings with these human-panel
averages. We also compare individual human reviewers with the same averages
using the same metrics. Their agreement provides the human baseline for
interpreting AI performance. For score assessment, we report Mean Absolute Error (MAE), bias, Spearman's $\rho$, and quadratic weighted Cohen's $\kappa$. MAE measures the average absolute difference between an AI score
and the corresponding human-panel mean, with lower values indicating
closer numerical agreement. Bias is the mean signed difference
between AI and human-panel scores: positive values indicate
over-scoring relative to humans, negative values indicate
under-scoring, and values closer to zero indicate less systematic
deviation. Spearman's $\rho$ measures whether proposals receiving
higher human scores also tend to receive higher AI scores, with
higher values indicating stronger agreement in ordering.
Quadratic weighted Cohen's $\kappa$ measures agreement beyond
chance between AI scores and rounded human-panel scores,
penalizing larger score disagreements more heavily; higher
values indicate stronger agreement.
For ranking, we report Kendall's $\tau$, Spearman's $\rho$, rank MAE, top-1 accuracy, bottom-1 accuracy, and pairwise concordance. Top-1 accuracy is the percentage of research-question groups
in which the AI and human panel identify the same strongest
proposal. Bottom-1 accuracy similarly measures agreement on
the weakest proposal. When an extreme rank is tied, a match
is counted if the AI and human sets share at least one proposal.
Pairwise concordance is the percentage of within-question
proposal pairs ordered in the same direction by the AI and
human panel, excluding pairs tied by either evaluator.
Higher values indicate stronger agreement for all three metrics.

\begin{table}[t]
\centering
  \caption{Score agreement with the human panel. The human reviewer–panel row compares individual reviewers’ scores with the panel mean for each proposal and criterion, with agreement measures averaged across criteria. AI rows compare each model's scores to the same panel mean. Lower MAE indicates closer score calibration; higher Spearman $\rho$ and $\kappa$ indicate stronger agreement. Bias and $\kappa$ are not reported for the human baseline.}
  \label{tab:rq2-score-summary}
  \small
  \begin{tabular}{lrrrr}
    \toprule
    Judge/model & MAE ($\downarrow$) & Bias ($\downarrow$) & Spearman  $\rho$ ($\uparrow$) & $\kappa$ ($\uparrow$) \\
    \midrule
    Individual human reviewer vs.\ panel mean & 0.57 & -- & \textbf{0.58} & -- \\
    \midrule
    GPT-4.1              & \textbf{0.48} &  0.23 & 0.42 & \textbf{0.35} \\
    Gemini~2.5 Flash     & 0.74 &  0.03 & 0.39 & 0.29 \\
    Claude Sonnet~4.6    & 0.88 & $\textbf{-0.83}$ & 0.46 & 0.20 \\
    Qwen~2.5 72B         & 0.58 &  0.34 & 0.28 & 0.18 \\
    \bottomrule
  \end{tabular}
\end{table}
\begin{figure}[t]
  \centering
  \includegraphics[width=\linewidth]{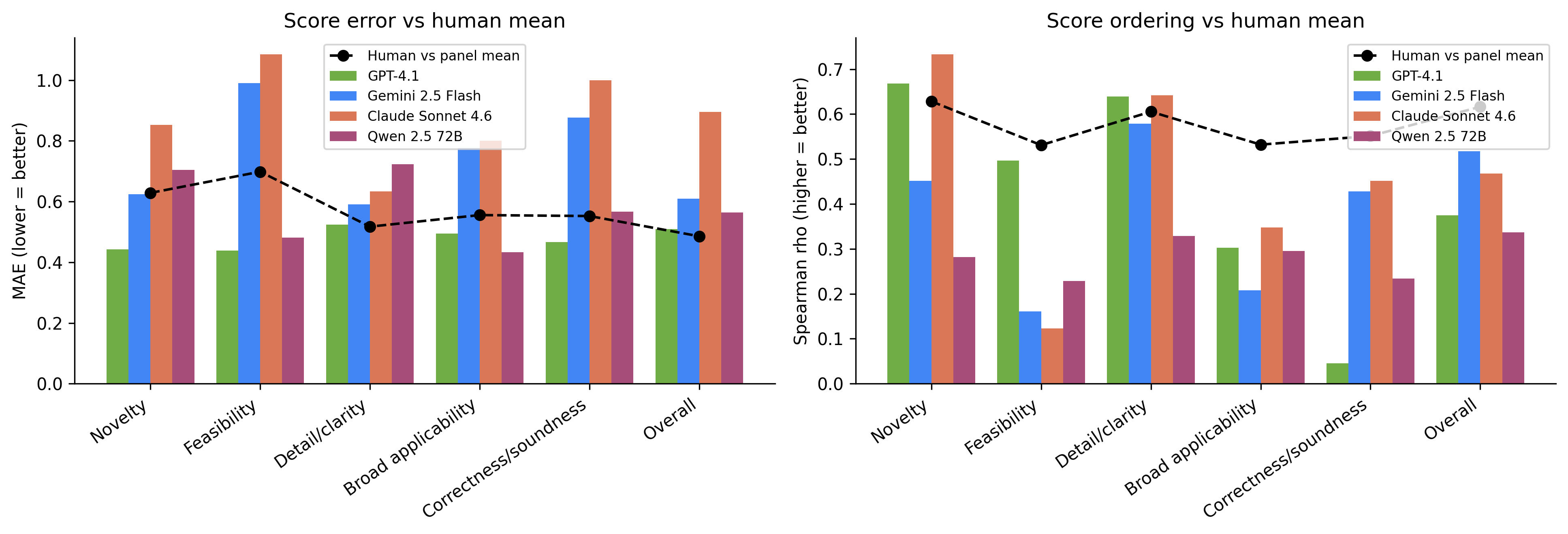}
    \vspace{-10mm}
  \caption{Criterion-specific AI vs. human panel (left: MAE ($\downarrow$); right: Spearman ($\uparrow$)).}
  \label{fig:rq2-score-dimensions}
    \vspace{-3mm}
\end{figure}

\paragraph{\textbf{Results.}}
Table~\ref{tab:rq2-score-summary} compares AI judges against the human panel, using individual human reviewers as a reference baseline.
GPT-4.1 achieves the best score calibration among AI models, with the lowest MAE (0.48), outperforming the human reviewer--panel baseline (0.57), and the highest weighted $\kappa$ (0.35).
However, its Spearman correlation (0.42) remains below the human baseline (0.58), indicating that numerical proximity does not guarantee consistent ordering of proposal quality.
Claude Sonnet achieves the highest score correlation among AI judges ($\rho=0.46$) but systematically underestimates quality (bias $=-0.83$).
Gemini exhibits the smallest bias (0.03), whereas Qwen achieves human-comparable MAE (0.58) but substantially weaker correlation ($\rho=0.28$).
These differences suggest that score calibration, systematic bias, and relative discrimination capture distinct aspects of evaluation reliability.

Figure~\ref{fig:rq2-score-dimensions} further breaks down the evaluation of Table \ref{tab:rq2-score-summary} into different scientific-quality criteria. It reveals that detail/clarity is the criterion on which AI judges most consistently approach human-level agreement in both MAE and Spearman correlation.
In contrast, novelty and feasibility exhibit larger gaps for most models, particularly in preserving human score orderings.
For broad applicability and correctness/soundness, GPT-4.1 and Qwen achieve relatively low MAE despite substantially weaker correlations with human judgments.
This suggests that AI judges may reproduce average human scoring patterns without consistently distinguishing stronger from weaker proposals, particularly on criteria requiring broader scientific judgment.

Table~\ref{tab:rq2-rank-summary} additionally examines whether AI judges preserve human preferences when ranking proposals addressing the same research question.
Overall, AI judges exhibit weaker ranking agreement with the human panel than individual human reviewers, with substantial variation across models.
GPT-4.1 achieves the strongest overall ranking agreement among AI models, with Kendall's $\tau=0.42$, Spearman's $\rho=0.48$, and rank MAE of 0.61, compared with 0.57, 0.61, and 0.54 for the human baseline, respectively.
Its pairwise concordance (80.5\%) approaches the human baseline (83.1\%), although its top-1 and bottom-1 accuracies (55\% and 60\%) remain below human levels (70\% and 75\%).
Claude Sonnet achieves slightly higher top-1 accuracy (60\%) but weaker overall rank correlations, while Gemini and Qwen show substantially lower agreement across most ranking metrics.
Importantly, these ranking results contrast with the score agreement in Table~\ref{tab:rq2-score-summary}: GPT-4.1 achieves lower score MAE than the individual-human baseline (0.48 vs.\ 0.57), yet falls short of human-level ranking agreement.
The discrepancy is particularly pronounced for Qwen, whose score MAE (0.58) is comparable to the human baseline (0.57), but whose ranking correlations are near zero (Kendall's $\tau=0.02$, Spearman's $\rho=0.07$).
Together, these findings demonstrate that numerical agreement with human scores does not necessarily translate into faithful preservation of human preferences over scientific proposals.

\begin{table}[t]
  \caption{Rank agreement with the human panel within RQs, averaged across 20 RQs. The human reviewer--panel row compares individual human rankings to the panel mean rank; AI rows compare model ranks to the same panel mean rank. Higher Kendall $\tau$, Spearman $\rho$, top-1, bottom-1, and pairwise concordance indicate stronger alignment; lower rank MAE indicates closer rank calibration.}
  \label{tab:rq2-rank-summary}
  \small
  \resizebox{\linewidth}{!}{
  \begin{tabular}{lrrrrrr}
    \toprule
    Judge/model & Kendall $\tau$ & Spearman $\rho$ & Top-1 (\%) & Bottom-1 (\%) & Rank MAE & Pairwise concordance (\%) \\
    \midrule
    Human reviewer vs.\ panel mean & \textbf{0.57} & \textbf{0.61} & \textbf{70.0} & \textbf{75.0} & \textbf{0.54} & \textbf{83.1} \\
    \midrule
    GPT-4.1              & 0.42 & 0.48 & 55 & 60 & 0.61 & 80.5 \\
    Gemini~2.5 Flash     & 0.21 & 0.25 & 50 & 35 & 0.76 & 67.8 \\
    Claude Sonnet~4.6    & 0.37 & 0.39 & 60 & 60 & 0.80 & 71.3 \\
    Qwen~2.5 72B         & 0.02 & 0.07 & 30 & 40 & 0.87 & 59.8 \\
    \bottomrule
  \end{tabular}
  }
\end{table}

\paragraph{\textbf{Conclusions.}}
i) AI judges exhibit meaningful but incomplete alignment with human evaluation, with substantial variation across models, criteria, and evaluation objectives. GPT-4.1 shows the strongest overall agreement, approaching the individual-human baseline in score calibration and leading most AI models in ranking agreement, but a clear human--AI gap remains, particularly in preserving proposal ordering. ii) More importantly, our results reveal that \emph{score calibration does not imply preference alignment}: an AI judge can produce scores numerically close to the human panel mean while failing to preserve which proposals humans consider better, as observed for broad applicability and correctness/soundness. iii) Reliability is also criterion-dependent. AI judges are most human-like on detail/clarity, whereas more subjective or expertise-intensive criteria such as novelty and feasibility retain larger human--AI gaps. Taken together, these findings suggest that current AI judges are better positioned as \emph{evaluation assistants} than as general replacements for human reviewers: they can provide useful and sometimes human-comparable signals, but their reliability depends critically on the evaluation criterion and whether the downstream task requires calibrated scoring or faithful preference ranking.

\subsection{RQ3: Does AI score Human+AI proposals more favorably than Human+Human proposals?}
\label{sec:rq3}

\paragraph{\textbf{Motivation.}}
This question is motivated by a potential evaluator-side bias: if an AI judge systematically assigns higher scores to proposals written with AI assistance, then AI-as-judge evaluation may amplify the apparent benefit of AI collaboration.

\paragraph{\textbf{Method \& Evaluation Metrics.}}
We compare 36 Human+AI proposals and 34 Human+Human proposals.
As RQ2 has indicated GPT4 is the most reliable AI judge, here we report GPT4.
For both GPT-4.1 and the human panel, we analyze the average of all six ratings (five rubric dimension scores plus overall quality) and the overall quality score separately.

For each proposal $p$, let $AI_p$ indicate whether the proposal was produced in the Human+AI condition.
We first compare the pooled mean outcomes between Human+AI and Human+Human proposals:
\[
\Delta = \mathbb{E}[Y_p \mid AI_p=1] - \mathbb{E}[Y_p \mid AI_p=0].
\]
For each outcome, we report the Human+AI mean, Human+Human mean, mean difference, bootstrap 95\% confidence interval, Cohen's $d$, Cliff's $\delta$, Welch's $t$-test, and Mann--Whitney test.

We also compute dimension-level effects for each rubric dimension.
For each dimension $d$, we compare GPT-4.1 scores, human-panel scores, and GPT-minus-human bias:
\[
b_{p,d} = s^{\mathrm{GPT}}_{p,d} - \bar{s}_{p,d}.
\]
Positive values of $b_{p,d}$ indicate that GPT-4.1 scores a proposal higher than the human panel on that dimension.

Finally, we evaluate the likelihood of receiving high scores, where high score means a score of at least 4 on the 1--5 scale.
We compare the fraction of Human+AI and Human+Human proposals receiving high GPT-4.1 and human scores using Fisher's exact test.

\paragraph{\textbf{Results}.}
Table~\ref{tab:rq3-aggregate-summary} summarizes the dimension-aggregated comparison.
GPT-4.1 assigns higher scores to Human+AI proposals than to Human+Human proposals.
The GPT-4.1 mean score over all criteria is 3.90 for Human+AI proposals and 3.76 for Human+Human proposals, a difference of 0.13 points.
The bootstrap 95\% confidence interval is [0.01, 0.25], with Cohen's $d=0.51$.
In contrast, the human-panel mean score is identical across conditions: 3.605 for Human+AI and 3.605 for Human+Human, a difference of exactly $0.00$.

The same pattern appears for overall scores.
GPT-4.1 assigns Human+AI proposals a higher overall score than Human+Human proposals (3.94 versus 3.76; difference 0.18), while human overall scores differ only slightly (3.45 versus 3.41; difference 0.04).
The GPT-minus-human bias is also larger for Human+AI proposals: 0.29 versus 0.16 when averaged over all criteria.

\begin{table*}[t]
  \centering
  \caption{Aggregate score comparison by collaboration condition.
  Difference is Human+AI minus Human+Human. The mean score is
  calculated across all six ratings: the five rubric dimensions
  and overall quality. Confidence intervals are proposal-level
  bootstrap 95\% intervals.}
  \label{tab:rq3-aggregate-summary}
  \small
  \resizebox{\textwidth}{!}{%
  \begin{tabular}{lrrrrrrr}
    \toprule
    Outcome & Human+AI & Human+Human & Difference
      & 95\% CI & Cohen's $d$ & Welch $p$ \\
    \midrule
    GPT mean across six ratings
      & 3.898 & 3.765 & 0.133
      & $[0.013,\,0.253]$ & 0.514 & 0.036 \\
    Human mean across six ratings
      & 3.605 & 3.605 & 0.000
      & $[-0.170,\,0.170]$ & 0.001 & 0.997 \\
    $\Delta$(GPT-human)
      & 0.293 & 0.160 & 0.133
      & $[-0.000,\,0.269]$ & 0.459 & 0.059 \\
    \midrule
    GPT overall-quality score
      & 3.944 & 3.765 & 0.180
      & $[0.033,\,0.353]$ & 0.524 & 0.036 \\
    Human overall-quality score
      & 3.454 & 3.412 & 0.042
      & $[-0.211,\,0.289]$ & 0.077 & 0.747 \\
    $\Delta$(GPT-human)
      & 0.491 & 0.353 & 0.138
      & $[-0.104,\,0.389]$ & 0.259 & 0.279 \\
    \bottomrule
  \end{tabular}%
  }
\end{table*}

\textbf{We next examine whether the two collaboration types (Human+AI vs.\ Human+Human) differ in their likelihood of receiving high scores.} As Table \ref{tab:rq3-high-score} lists, GPT-4.1 is also more likely to assign high scores to Human+AI proposals.
41.7\% of Human+AI proposals have GPT-4.1 mean scores of at least 4, compared with 23.5\% of Human+Human proposals.
For GPT-4.1 overall scores, the difference is larger in absolute prevalence: 94.4\% of Human+AI proposals receive an overall score of at least 4, compared with 76.5\% of Human+Human proposals (Fisher's exact $p=0.043$). In contrast, the human-panel high-score rates do not show the same pattern.
For the human mean score over all criteria, 16.7\% of Human+AI proposals and 17.6\% of Human+Human proposals reach the threshold.
For human overall scores, the rates are 27.8\% and 20.6\%, respectively.

\begin{table}[t]
  \centering
  \caption{High-score rates by collaboration condition. High score means score $\geq 4$.}
  \label{tab:rq3-high-score}
  \small
  \begin{tabular}{lrrrr}
    \toprule
    Outcome & Human+AI & Human+Human & Difference & Fisher $p$ \\
    \midrule
    GPT mean $\geq 4$     & 41.7\% & 23.5\% & 18.1 pp & 0.131 \\
    GPT overall $\geq 4$  & 94.4\% & 76.5\% & 18.0 pp & 0.043 \\
    Human mean $\geq 4$   & 16.7\% & 17.6\% & $-$1.0 pp & 1.000 \\
    Human overall $\geq 4$& 27.8\% & 20.6\% &  7.2 pp & 0.582 \\
    \bottomrule
  \end{tabular}
\end{table}

\begin{figure}[t]
  \centering
  \includegraphics[width=\linewidth]{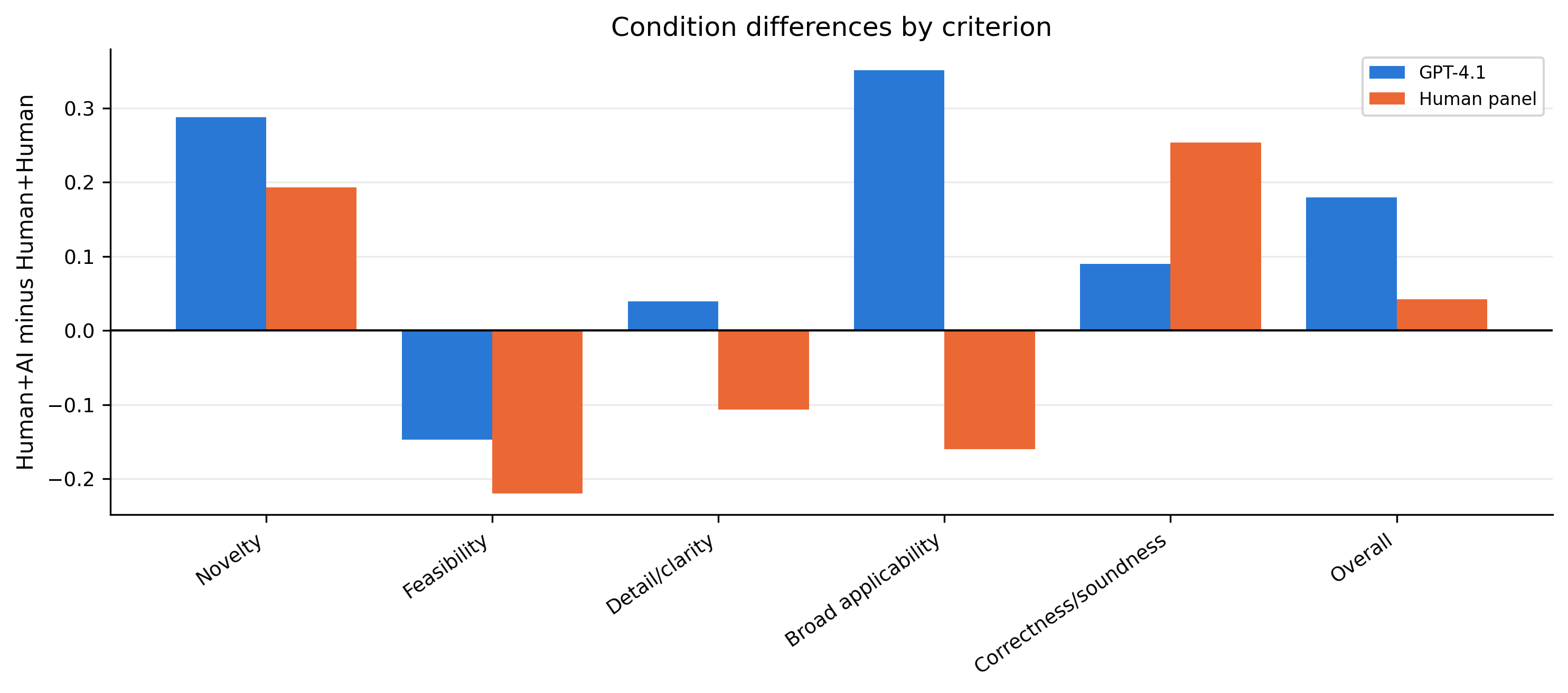}
  \vspace{-10mm}
  \caption{Condition differences by dimension. Bars report Human+AI minus Human+Human differences for GPT-4.1 scores and human-panel scores.}
  \label{fig:rq3-dimension-differences}
    \vspace{-5mm}
\end{figure}

\textbf{Furthermore, we examine whether this pattern varies across evaluation dimensions.} The largest dimension-level GPT-4.1 difference appears for broad applicability. As Figure \ref{fig:rq3-dimension-differences} illustrates, 
GPT-4.1 scores Human+AI proposals 0.35 points higher than Human+Human proposals on broad applicability, while the human panel scores Human+AI proposals 0.16 points lower on the same dimension.
Thus, the GPT-minus-human bias gap is largest for broad applicability: 0.31 for Human+AI proposals versus $-0.21$ for Human+Human proposals, a difference of 0.51 points. Other dimensions show smaller or mixed patterns.
GPT-4.1 gives Human+AI proposals higher novelty scores by 0.29 points, while the human-panel novelty difference is 0.19 points.
For feasibility, both GPT-4.1 and humans score Human+AI proposals lower than Human+Human proposals, but humans show the larger negative difference.
For correctness/soundness, the human panel scores Human+AI proposals higher, while the GPT-minus-human bias is actually smaller for Human+AI proposals.

\textbf{Finally, does this pattern persist after accounting for human scores and evaluation dimensions?}
Regression analyses confirm that the observed pattern is not driven solely by differences in human-assessed proposal quality or criterion composition. At the proposal level, Human+AI is associated with a 0.13-point increase in GPT-4.1's mean score ($p=0.035$), whereas no corresponding difference appears in the human panel ($\beta=0.00$, $p=0.997$). The resulting increase in the GPT--human score gap is also 0.13 points, although this proposal-level test is marginal ($p=0.059$).

When the six evaluation criteria are modeled jointly with criterion fixed effects, the Human+AI effect remains 0.13 for both GPT-4.1 scores ($p=0.002$) and the GPT--human discrepancy ($p=0.011$). These results provide additional evidence that GPT-4.1's more favorable evaluation of Human+AI proposals persists after accounting for criterion-level variation, although the evidence for the evaluator discrepancy is stronger in the dimension-stacked analysis than at the proposal level.

\noindent\textbf{Conclusions.}
i) GPT-4.1 exhibits a systematic tendency to favor Human+AI proposals that is
not mirrored by human reviewers. It assigns Human+AI proposals higher mean
and overall scores and is more likely to place them in the high-score range,
whereas the human panel shows little or no corresponding advantage.
ii) This evaluator discrepancy is \emph{dimension-dependent rather than
uniform}: it is most pronounced for broad applicability, where GPT-4.1 favors
Human+AI proposals while human reviewers show the opposite preference.
iii) The pattern persists after accounting for human scores and criterion
differences, although the proposal-level GPT--human discrepancy is only
marginally significant. Overall, these results reveal an
\emph{evaluator-side amplification}: an AI judge can amplify the apparent
benefit of Human+AI collaboration even when that benefit is not supported by
human evaluation. This highlights that aggregate human--AI agreement alone
is insufficient for validating AI-as-judge systems; agreement should also be
examined across the experimental conditions being compared.

\subsection{RQ4: Can We Detect Whether a Proposal Was Produced by Human--AI or Human--Human Collaboration?}
\label{sec:rq4}

\paragraph{\textbf{Motivation.}}
As LLMs become increasingly integrated into scientific workflows, AI participation in research ideation is likely to become increasingly common. This raises an important question: \emph{can we determine whether a research proposal was developed through Human--AI or Human--Human collaboration?} Such attribution is important for academic integrity, authorship credit, and transparency regarding the respective contributions of humans and AI to scientific work. Beyond attribution, distinguishability also provides a lens into the capability of AI for research ideation. If Human--AI proposals are difficult to distinguish from Human--Human proposals, this may indicate substantial overlap between methodologies developed with and without AI assistance. We therefore investigate whether the collaboration condition---Human--AI versus Human--Human---can be identified from the proposed research methodology.

\paragraph{\textbf{Method}.}
The input consists of the research question and proposal methodology.
The model returns exactly one of two labels: Human+Human or Human+AI.

\paragraph{Prompting.} We first evaluate pretrained Qwen3-14B and GPT-4.1 without task-specific fine-tuning or labeled examples as zero-shot baselines.
Both receive the same binary-classification instruction and the same research-question and methodology fields.
GPT-4.1 returns a schema-constrained label at temperature zero, while Qwen selects between the two labels using their conditional log-likelihoods.

\paragraph{Finetuned classifier.} We then train the Qwen3-14B classifier by supervised fine-tuning on the synthetic training data using 4-bit QLoRA~\cite{dettmers2023qlora}.
The LoRA adapters use rank 16, scaling factor 32, dropout 0.05, and target the attention and feed-forward projection layers.
Training runs for two epochs with learning rate $2\times10^{-4}$, effective batch size 8 through gradient accumulation, cosine scheduling, and a maximum sequence length of 2,048 tokens.

\paragraph{Simple baselines.}
We include uniform random guessing, which predicts either
collaboration label with equal probability, and a majority-class
reference, which predicts Human+AI for every proposal.
Random guessing has an expected accuracy of 50\%.
The majority-class reference achieves 51.4\% accuracy and
0.340 macro-F1 on the real evaluation set, which contains
36 Human+AI and 34 Human+Human proposals.
Both references apply to the unified and pre-unified
representations because their proposal identities and labels
are identical.

\paragraph{Evaluation Metrics.} We report accuracy and macro-F1.

\paragraph{\textbf{Results}}

\begin{figure}[t]
  \centering
  \includegraphics[width=\linewidth]{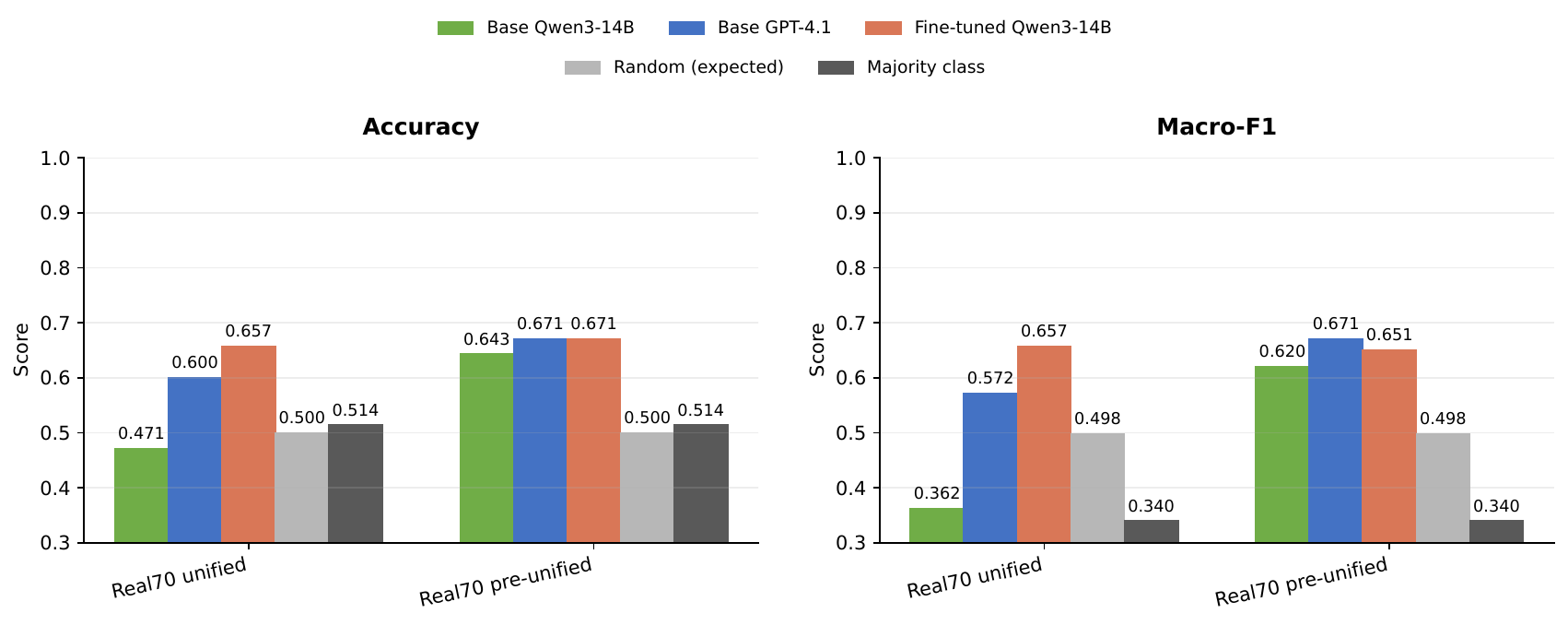}
    \vspace{-8mm}
  \caption{Accuracy and macro-F1 on the same 70 real proposals
  using unified and pre-unified text. Qwen3-14B is evaluated
  zero-shot and after synthetic-data fine-tuning; GPT-4.1 is
  evaluated zero-shot. Random denotes the exact expected
  performance of independent, equal-probability label guesses.
  The majority-class reference always predicts Human+AI,
  the more frequent label in this evaluation set
  (36 of 70 proposals). Baselines apply to both text
  representations. Both vertical axes start at 0.3.}
  \label{fig:rq4-performance}
    \vspace{-5mm}
\end{figure}

Figure~\ref{fig:rq4-performance} summarizes the results. 
The primary result is transfer to the 70 real proposals.
On unified proposals, zero-shot Qwen reaches 47.1\% accuracy, GPT-4.1 reaches 60.0\%, and fine-tuned Qwen reaches 65.7\%.
Relative to zero-shot Qwen, fine-tuning improves unified accuracy by 18.6 percentage points and macro-F1 from 0.362 to 0.657.
On pre-unified proposals, zero-shot Qwen reaches 64.3\% accuracy, while GPT-4.1 and fine-tuned Qwen both reach 67.1\%.
GPT-4.1 has the strongest pre-unified macro-F1 (0.671), indicating more balanced class performance than fine-tuned Qwen (0.651).
Fine-tuned Qwen therefore performs most consistently across unified and pre-unified representations, whereas GPT-4.1 is competitive specifically on pre-unified text.

\paragraph{\textbf{Conclusions.}}
i) Proposal methodologies contain detectable signals of their underlying
collaboration condition: both GPT-4.1 and fine-tuned Qwen perform above chance
on real proposals, with the best models reaching approximately 66--67\%
accuracy. ii) Synthetic supervision improves Qwen's ability to
detect the collaboration condition, particularly on unified proposals,
suggesting that collaboration-related patterns learned from synthetic data can
partially transfer to real research proposals. However, the moderate accuracy on real proposals indicates that
these learned patterns are insufficient for reliable
classification of individual proposals.
iii) Overall, collaboration provenance is \emph{detectable but far from
reliably identifiable}: the moderate classification accuracy suggests that
Human--Human and Human--AI research methodologies exhibit systematic differences,
while their considerable overlap makes dependable attribution at the
individual-proposal level difficult. Thus, methodology alone provides useful
evidence of collaboration provenance, but is insufficient for reliable attribution
of an individual proposal's collaboration condition.

\subsection{RQ5: How Does Human--AI Collaboration Change Participants' Perceptions of Generative AI?}
\label{sec:rq5}

\paragraph{\textbf{Motivation.}}
As generative AI becomes increasingly involved in scientific research, its impact depends not only on what AI can objectively accomplish, but also on how researchers perceive its capabilities and limitations. Importantly, such perceptions may be formed before researchers have substantial experience collaborating with AI and may therefore be miscalibrated. Direct Human--AI collaboration provides an opportunity to test whether hands-on experience changes these beliefs: does working with GenAI strengthen researchers' confidence in its capabilities, reveal limitations that temper their expectations, or reshape which aspects of research they consider AI well suited for? Understanding this perception shift provides insight into how actual collaboration experience calibrates researchers' trust in and expectations of AI, with implications for its future adoption and appropriate use in scientific workflows. 

\paragraph{\textbf{Method.}} We therefore compare participants' pre-study expectations with their post-collaboration perceptions across eight matched measures of GenAI capability and utility. 

The eight measures comprise 24 matched pre/post questionnaire
items. Table~\ref{tab:rq5-questionnaire-items} summarizes their
wording, item coverage, and response scales. Pre-study items
capture expectations or initial perceptions; post-study items
capture perceptions after the study. Response categories are
listed in ascending order of their harmonized 1--5 coding.

\begin{table*}[t]
  \centering
  \caption{Questionnaire measures used in the pre/post analysis.
  Question wording and item descriptions are summarized.
  Each matrix subitem counts as one item.}
  \label{tab:rq5-questionnaire-items}
  \footnotesize
  \renewcommand{\arraystretch}{1.2}
  \begin{tabularx}{\textwidth}{
    @{}
    p{0.15\textwidth}
    X
    p{0.27\textwidth}
    @{}
  }
    \toprule
    \textbf{Measure}
    & \textbf{Question summary and scope}
    & \textbf{Response options: 1 to 5} \\
    \midrule

    Confidence\newline
    (1 item)
    &
    \textbf{Pre:} How confident are you in integrating GenAI
    into your research workflow?
    \textbf{Post:} How confident are you now in integrating
    AI into your research-design workflow?
    &
    \textbf{Pre:} Not confident at all; Slightly confident;
    Neutral; Confident; Very confident.
    \textbf{Post:} Not at all; Slightly; Moderately;
    Mostly; Completely.
    \\

    \addlinespace
    Overall attitude\newline
    (1 item)
    &
    \textbf{Pre:} What is your overall attitude toward GenAI
    in research design before the study?
    \textbf{Post:} What is your current attitude toward GenAI
    in research design?
    &
    Very negative; Somewhat negative; Neutral;
    Somewhat positive; Very positive.
    \\

    \addlinespace
    Helpfulness\newline
    (6 items)
    &
    \textbf{Pre/Post:} How helpful do you expect GenAI to be /
    perceive GenAI to be after the study?
    Items cover hypotheses and conceptual frameworks;
    methodology selection; literature synthesis;
    experimental or survey design; identifying biases and
    limitations; and writing assistance.
    &
    Not Helpful; Slightly Helpful; Moderately Helpful;
    Very Helpful; Extremely Helpful.
    \\

    \addlinespace
    Challenge severity\newline
    (7 items)
    &
    \textbf{Pre/Post:} To what extent do you expect /
    now perceive the following GenAI challenges?
    Items cover limited originality; limited depth and
    nuanced reasoning; flawed or impractical hypotheses;
    methodologies inconsistent with best practices;
    limited discipline-specific context; over-reliance
    weakening critical thinking and expertise; and ethical
    concerns, including bias and hallucinated sources.
    &
    Not a Challenge; Minor Challenge; Moderate Challenge;
    Significant Challenge; Major Challenge.
    \\

    \addlinespace
    Speed-up\newline
    (1 item)
    &
    \textbf{Pre:} How much do you expect GenAI to speed up
    research ideation?
    \textbf{Post:} How much do you think GenAI actually
    speeds up research ideation?
    Examples include brainstorming, generating hypotheses,
    and designing methodology.
    &
    No noticeable impact; Slightly faster; Moderately faster;
    Significantly faster; Extremely fast.
    \\

    \addlinespace
    Human effort\newline
    (1 item)
    &
    \textbf{Pre/Post:} How much human effort do you anticipate /
    did it actually take to refine GenAI outputs?
    Examples include prompt adjustments, multi-turn interaction,
    fact-checking, improving coherence and rigor, alignment
    with research goals, citation, and ethical standards.
    &
    Minimal effort (little or no modification);
    Slight effort (minor edits);
    Moderate effort (significant modifications);
    Considerable effort (substantial revision and validation);
    High effort (extensive rewriting).
    \\

    \addlinespace
    Capability enhancement\newline
    (6 items)
    &
    \textbf{Pre/Post:} To what extent do you believe /
    now perceive that GenAI enhances human research-design
    capabilities?
    Items cover compiling prior research and citations;
    generating novel ideas; synthesizing literature;
    improving systematic reviews and identifying gaps;
    critiquing methodologies; and structuring or refining
    proposals and experimental designs.
    &
    No Enhancement; Minor Enhancement; Moderate Enhancement;
    Significant Enhancement; Major Enhancement.
    \\

    \addlinespace
    Trust\newline
    (1 item)
    &
    \textbf{Pre:} How much do you trust GenAI's insights and
    recommendations for research design?
    \textbf{Post:} How much do you now trust AI-generated
    research-design suggestions?
    &
    Not at all; Slightly; Moderately; Mostly; Completely.
    \\

    \bottomrule
  \end{tabularx}
\end{table*}

For example, a higher attitude score means a more positive attitude, a higher challenge score means that challenges were perceived as more severe, and a higher human-effort score means that more effort was expected or required.

The pre-study questionnaire included responses from 47 unique participants, and the post-study questionnaire included responses from 45 unique participants.
Participant identifiers were reconciled across the surveys and collaboration-condition records.
Of the 55 unique participants overall, 19 were assigned to Human+AI, 31 to Human+Human, and five could not be reliably assigned to a collaboration condition and were therefore excluded from the analysis.
Table~\ref{tab:rq5-coverage} reports the complete coverage and mapping counts.

\begin{figure*}[t]
  \centering
  \includegraphics[width=\textwidth]{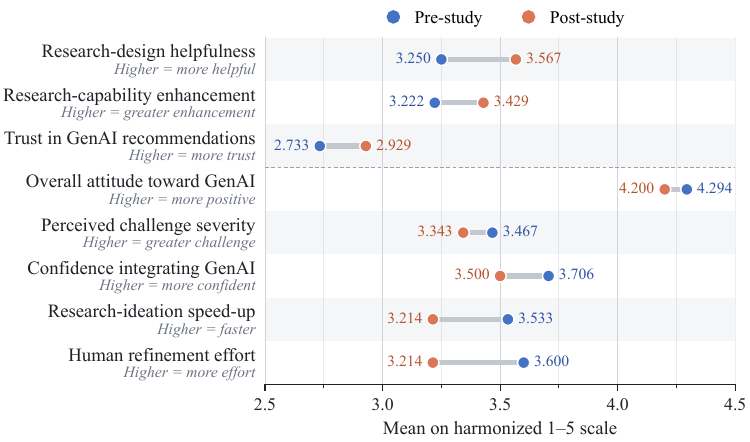}
    \vspace{-3mm}
  \caption{Human+AI participants' pre-study expectations and post-study perceptions using all available responses. Labels state the direction of each scale so that movement can be interpreted from the original survey wording. Lines show descriptive differences between group means and do not imply paired observations.}
  \label{fig:rq5-human-ai-prepost}
\end{figure*}

\paragraph{\textbf{Results.}}
Figure~\ref{fig:rq5-human-ai-prepost} reports the Human+AI pre/post comparison.
The largest positive descriptive change was in perceived helpfulness across six research-design tasks, which increased from 3.250 to 3.567 ($+0.317$).
Perceived enhancement of research-design capabilities increased from 3.222 to 3.429 ($+0.206$), and trust in GenAI recommendations increased from 2.733 to 2.929 ($+0.195$).

The largest negative descriptive changes concerned expected versus experienced effort and speed.
The score for human effort needed to refine GenAI outputs decreased from 3.600 to 3.214 ($-0.386$), indicating that participants reported less required effort after the study than they had expected beforehand.
The GenAI speed-up score decreased from 3.533 to 3.214 ($-0.319$), indicating that the experienced speed-up was lower than initially expected.
Perceived challenge severity also decreased slightly, from 3.467 to 3.343 ($-0.124$).

Overall attitude positivity remained high but decreased slightly from 4.294 to 4.200 ($-0.094$).
Confidence in integrating GenAI decreased from 3.706 to 3.500 ($-0.206$).

\paragraph{\textbf{Conclusions.}}
i) Direct Human--AI collaboration is associated with a more differentiated,
rather than uniformly more positive, perception of GenAI. Participants
reported greater task-specific helpfulness, capability enhancement, and trust,
while also finding that less human refinement effort was required and that
GenAI-related challenges were less severe than initially expected.
ii) Experience also corrected some optimistic expectations: the realized
speed-up was lower than anticipated, while confidence in integrating GenAI and
overall attitude positivity decreased slightly.
iii) Taken together, these shifts suggest a \emph{recalibration of expectations
through experience}: participants came to recognize greater practical utility
from GenAI while simultaneously developing more tempered expectations about
its efficiency and their ability to integrate it into research workflows.
Thus, Human--AI collaboration appears to reshape researchers' mental models of
GenAI in multiple directions rather than simply increasing or decreasing
their overall enthusiasm for it. As these results compare available pre- and
post-study responses rather than fully paired observations, they should be
interpreted as descriptive evidence of perception recalibration rather than
causal within-participant change.

\clearpage

\section{Conclusion}

As AI increasingly participates in both creating and evaluating scientific ideas, understanding its influence on scientific discovery becomes essential. Our findings reveal that AI collaboration does not necessarily improve research quality beyond human collaboration, while iterative refinement remains a powerful driver of improvement. More importantly, AI judges can systematically favor AI-assisted proposals despite comparable human evaluations, raising concerns about feedback loops in AI-mediated science. Meanwhile, collaboration leaves only weakly detectable traces in scientific artifacts and reshapes researchers' perceptions of AI's capabilities and limitations. Together, these findings suggest that the future of AI-driven science depends not simply on making AI a more capable collaborator, but on understanding and governing the interactions among human creativity, AI assistance, and scientific evaluation. Ultimately, trustworthy AI for science must advance not only how knowledge is created, but also how its quality, provenance, and value are established.

\bibliography{custom}
\bibliographystyle{iclr2026_conference}
\clearpage
\appendix
\section{RQ1 Supplementary Figures}
\label{app:rq1-supplementary}

\begin{figure}[htbp]
  \centering
  \includegraphics[
    width=0.8\linewidth,
    height=0.35\textheight,
    keepaspectratio
  ]{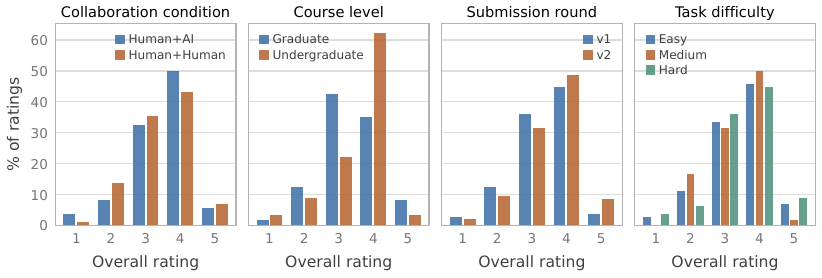}
  \caption{Proposal-level mean overall-rating distributions by
  collaboration condition, course level, submission round, and
  task difficulty. Points are individual proposals.}
  \label{fig:rq1-overall-distributions}

  \vspace{1em}

  \includegraphics[
    width=0.8\linewidth,
    height=0.35\textheight,
    keepaspectratio
  ]{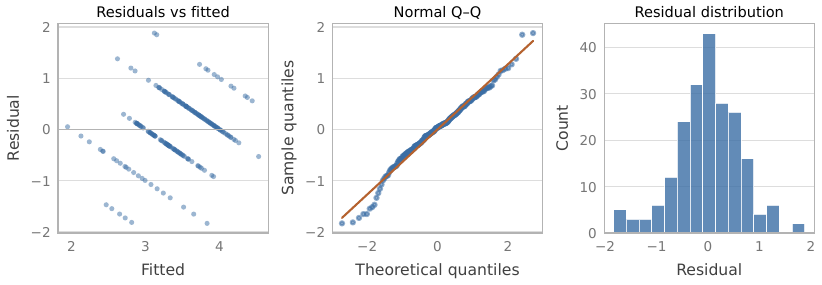}
  \caption{Diagnostics for the primary mixed-effects model of
  overall human ratings: residuals versus fitted values,
  normal Q--Q plot, residual distribution, and observed
  versus fitted ratings.}
  \label{fig:rq1-mixed-diagnostics}
\end{figure}

Figure~\ref{fig:rq1-overall-distributions} shows proposal-level overall-rating
distributions for each study attribute. Figure~\ref{fig:rq1-mixed-diagnostics}
reports diagnostics for the primary mixed-effects model. The visible residual
bands follow from the discrete 1--5 outcome; the Q--Q plot additionally shows
departures in the lower tail.

\clearpage
\section{RQ2 Confusion Matrices}
\label{app:rq2-confusion}

\begin{figure}[htbp]
  \centering
  \includegraphics[
    width=0.9\linewidth,
    height=0.35\textheight,
    keepaspectratio
  ]{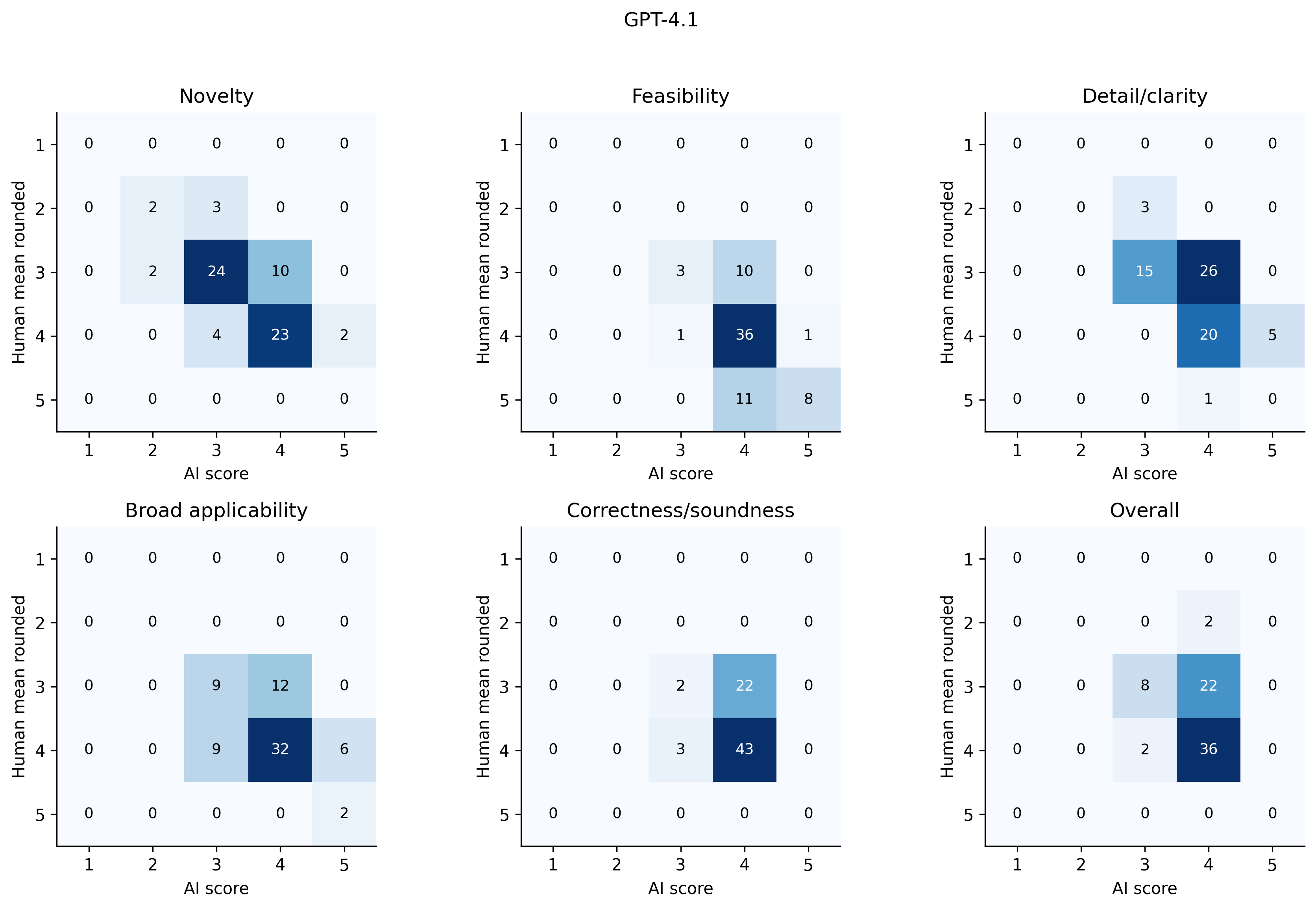}
  \caption{Confusion matrices for GPT-4.1 scores versus
  rounded mean human scores.}
  \label{fig:confusion-gpt41}

  \vspace{1em}

  \includegraphics[
    width=0.9\linewidth,
    height=0.35\textheight,
    keepaspectratio
  ]{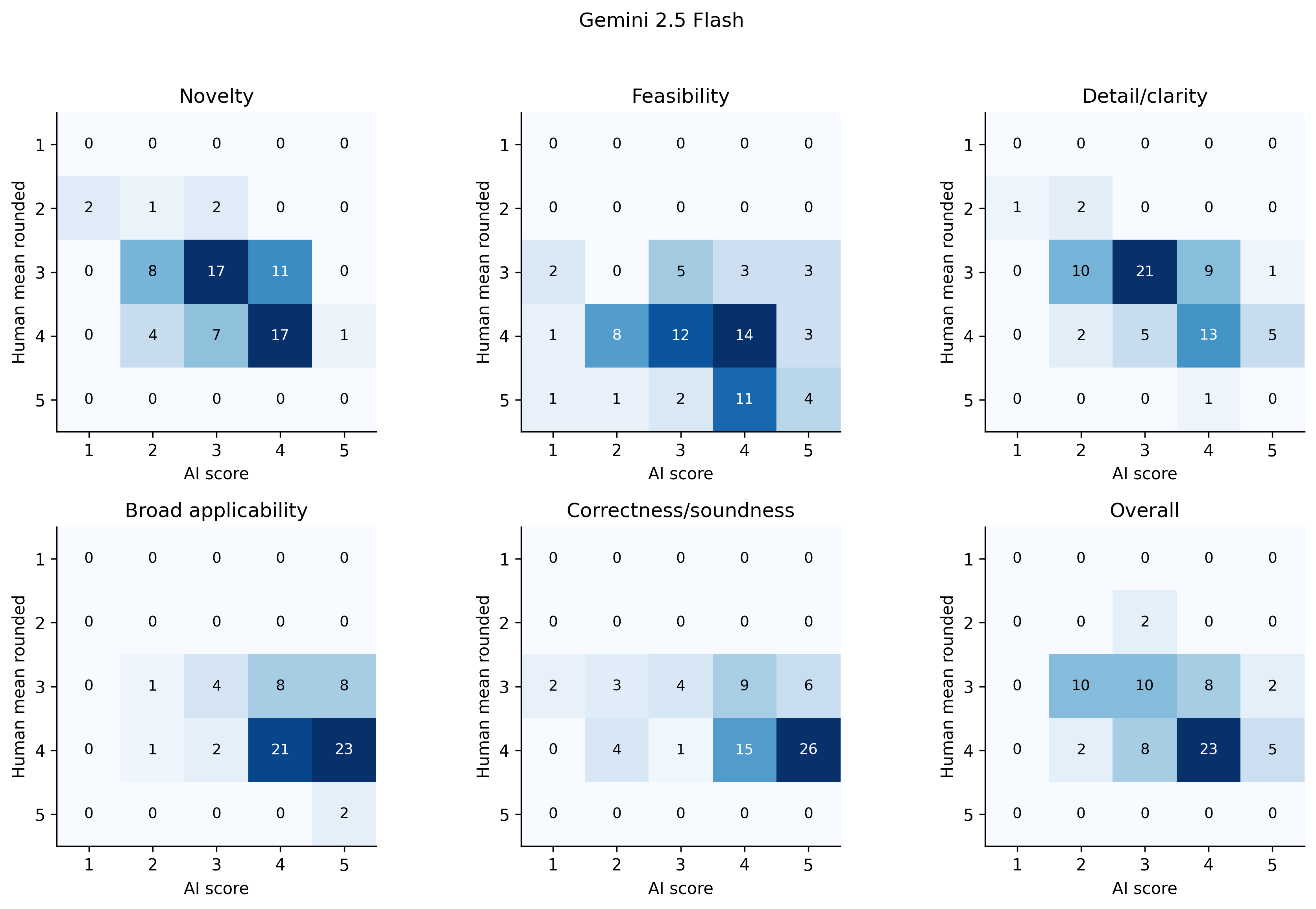}
  \caption{Confusion matrices for Gemini~2.5 Flash scores
  versus rounded mean human scores.}
  \label{fig:confusion-gemini}
\end{figure}

\begin{figure}[p]
  \centering
  \includegraphics[
    width=0.9\linewidth,
    height=0.35\textheight,
    keepaspectratio
  ]{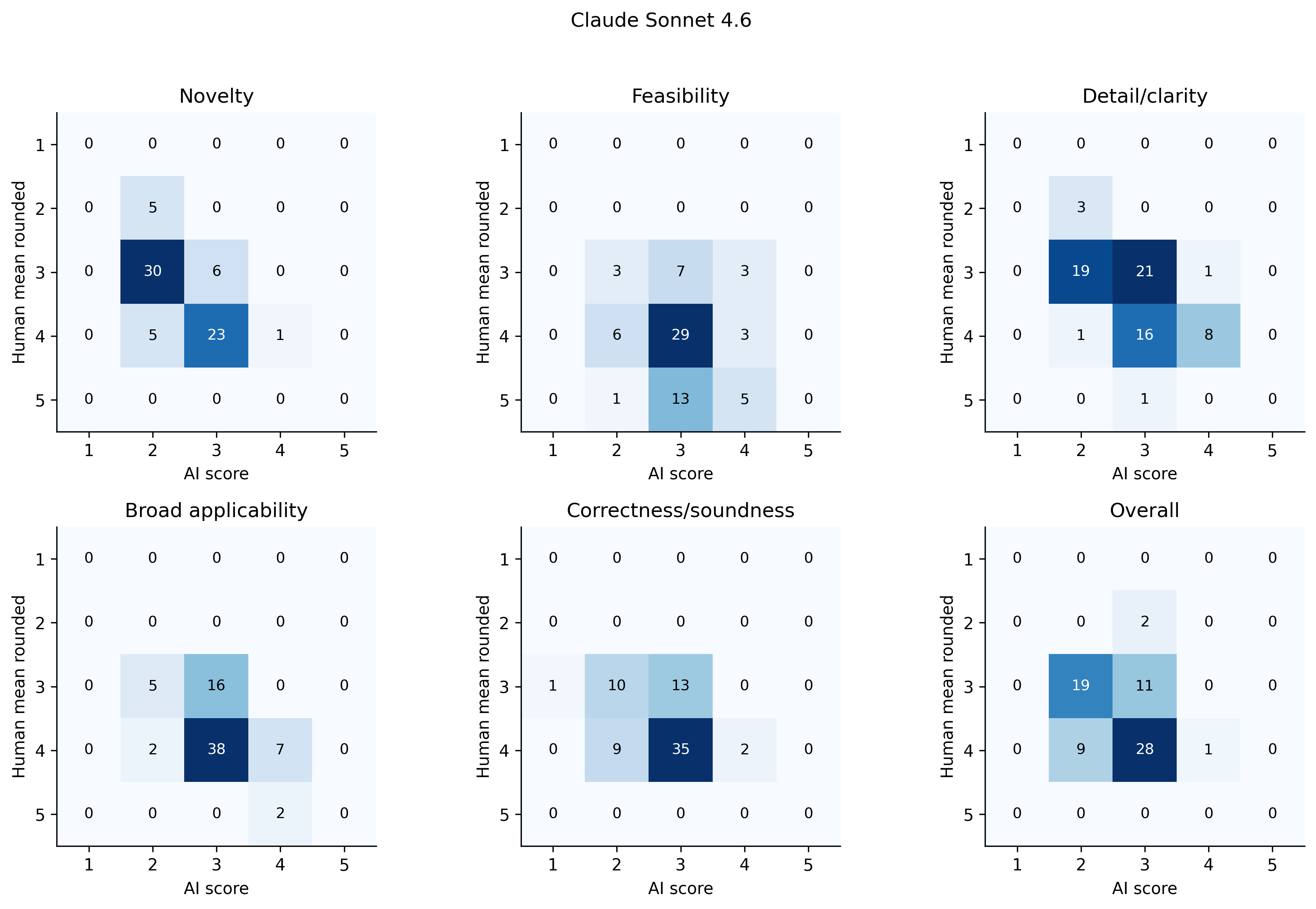}
  \caption{Confusion matrices for Claude Sonnet~4.6 scores
  versus rounded mean human scores.}
  \label{fig:confusion-claude}

  \vspace{1em}

  \includegraphics[
    width=0.9\linewidth,
    height=0.35\textheight,
    keepaspectratio
  ]{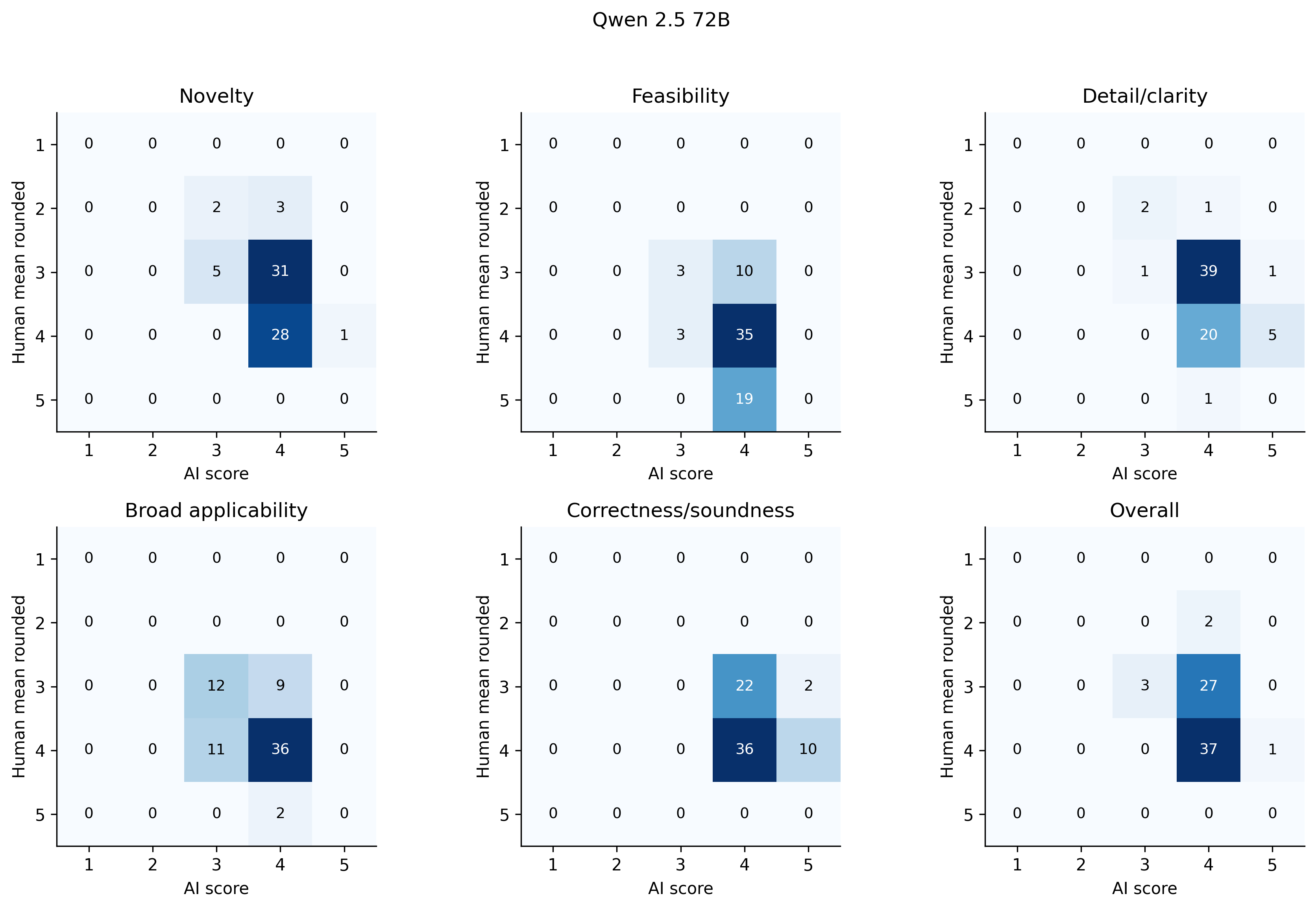}
  \caption{Confusion matrices for Qwen~2.5 72B scores
  versus rounded mean human scores.}
  \label{fig:confusion-qwen72b}
\end{figure}

Figures~\ref{fig:confusion-gpt41}--\ref{fig:confusion-qwen72b} show confusion matrices comparing each model's integer scores to rounded mean human scores across the six criteria.

\clearpage
\section{Research Questions Used in the Real-World Study}
\label{app:proposal-prompts}

Table~\ref{tab:proposal-prompts} lists the 20 NLP/ML research questions
assigned to the participating teams. 

\begin{table}[htbp]
  \centering
  \caption{Research questions addressed by the 70 real-world proposals.}
  \label{tab:proposal-prompts}
  \footnotesize
  \renewcommand{\arraystretch}{1.08}

  \begin{tabularx}{\textwidth}{@{}p{0.06\textwidth}X@{}}
    \toprule
    \textbf{ID} & \textbf{Research prompt} \\
    \midrule

    1 & How can an AI system generate novel research ideas in NLP? \\

    2 & Given a paper draft containing equations, how can a classifier determine whether an equation is correct in context? \\

    3 & How can a textual-entailment classifier be built for the biology domain without human-annotated, biology-specific training data? \\

    4 & How can LLM hallucinations be mitigated? \\

    5 & Build a model to predict house prices. \\

    6 & How well do authors of accepted conference papers address reviewers' required changes in their camera-ready versions? \\

    7 & A company stores consumer information across databases in which names, phone numbers, addresses, and other fields may be expressed differently or be missing. How can records referring to the same person be identified? \\

    8 & How can a classifier detect whether a student's report was written by AI or by the student? \\

    9 & Build a part-of-speech tagger that works across a variety of endangered languages. \\

    10 & Predict a paper's citation count over the next five years. \\

    11 & Given an LLM's output response, is it possible to recover the prompt that produced it? Propose a solution. \\

    12 & How can one detect whether a piece of text was used to pretrain an LLM? \\

    13 & Can LLMs identify when a question has no valid answer? \\

    14 & How can the novelty of a newly submitted paper be assessed? \\

    15 & How can each sentence in a student's report be classified as AI-written or student-written? \\

    16 & How can AI or LLMs be used to generate higher-quality paper reviews? \\

    17 & Detect conflicting claims in scientific papers in biology. \\

    18 & Propose a paper-search algorithm that retrieves papers from a user's desired information rather than only from user-provided keywords. \\

    19 & Build a classifier that identifies the native language of an essay's author. \\

    20 & Propose an algorithm to predict whether a student will pass or fail a course. \\

    \bottomrule
  \end{tabularx}
\end{table}

\end{document}